\def\im{\hbox{Im}\,}
\def\re{\hbox{Re}\,}

\def\ket#1{\left|\, #1\right\rangle}
\def\bra#1{\left\langle #1\right|}
\def\bracket#1#2{\left.\left\langle #1\right|#2\right\rangle}
\def\melement#1#2#3{\left.\left.\left\langle #1\right|#2\right|#3\right\rangle}

\documentclass[prl,twocolumn,showpacs]{revtex4}
\usepackage{hyperref}
\usepackage{amsfonts}

\begin{document}

\title{Amplification of Nonlocal Effects in Nonlinear Quantum Mechanics by Extreme Localization}
\author{George  Svetlichny}
\email{svetlich@mat.puc-rio.br}
\affiliation{Departamento de Matem\'atica, \\ Pontif{\'{\i}}cia
Universidade Cat{\'o}lica, \\ Rio de Janeiro, RJ,  Brazil}


\begin{abstract}
Due to its connection to the diffeomorphism group, nonlinear quantum mechanics may play an important role in quantum geometry. The Doebner-Goldin nonlinearity (arising from representations of the diffeomorphism group) amplifies nonlocal signaling effects under extreme localization, suggesting that even if greatly suppressed at low energies, such effects may be significant at the Planck scale.  This offers new perspectives on Planck-scale physics.
\end{abstract}

\maketitle

\section{Introduction}

The main objection to the various proposals for nonlinear quantum mechanics is an apparent conflict with causality\cite{polcal}. Although this issue was extensively discussed under many guises in the literature \cite{bonaal}, with various attempts at circumventing the problem, no truly relativistic causal nonlinear quantum mechanics has yet been created. Other seemingly problematic aspects of nonlinear quantum mechanics also seem to be related to space-time structure\cite{svetal}.

We argue here that vice  can be virtue.  Experimental evidence shows that at the low energies of atomic levels, nonlinear effects are suppressed by a factor of at least \(10^{-20}\) in relation to the linear ones\cite{nlexp1al}. Pushing this further, suppose nonlinear effects can only be exposed at Planck energies. Since at this scale space-time itself is thought to acquire quantum behavior, making its causal and manifold structure ill defined, the seeming conflicts of nonlinearity with space-time may no longer be operative. All attempt therefore of circumventing these difficulties could in the end be irrelevant as there may be no problematic low-energy  consequences. We are now  faced with the question: given the large suppression at low energies, could nonlinear effects be significant at high ones? If so, what would be considered an acausal effect at low energies, could at high energies provide completely new phenomena relevant for quantum space-time and early-universe cosmology. We shall use the more neutral term {\em  nonlocal\/} instead of acausal for these phenomena, deeming causality to be a 
post-Planckian concern.

The nonlinear equation introduced by Doebner and Goldin (DG) \cite{do-gopla162397} is particularly interesting in this respect as it arises in connection with representations of the diffeomorphism group (more exactly with current algebras). This particular nonlinearity may thus be of special relevance in any form of quantum geometry. We show that under extreme localization of states, the nonlocal behavior of the DG nonlinearity is amplified to the extent that it may be of the same order of magnitude as effects due to the liner terms of the evolution, which dominate at low energies. If such effect do operate at the Planck scale, we would be forced to completely modify our present views of this regime and 
consider alternatives to such endeavors as M-theory and loop quantum gravity for which linearity of quantum mechanics at all scales is an essential ingredient.  It is probably significant that the nonlocal effects of various other types of nonlinearities do not suffer this amplification, making the connection of the DG equation to the diffeomorphism group even more significant.

\subsection{Nonlocal signals}
Our analysis assumes that time evolution in nonlinear but that the usual linear quantum observables still do describe measurement processes (other nonlinear ones may also be present), that instantaneous state collapse is still a valid description, and that the square modulus of the wave function still provides a probability density for position measurements. All this can be questioned but many proposals for nonlinear quantum mechanics do assume this, and it can be deemed reasonable at low energies when nonlinear effects (including the nonlocal ones) are greatly suppressed. One needs some argument though to justify deeming results of such an analysis relevant for the high-energy Planck regime. Many of the results of the measurement process are reproducible to a high degree of accuracy by decoherence. Decoherence should be an operative process for nonlinear evolution at all scales and we consider the analysis carried out below as an idealized sketch of a decoherence analysis. Quantitative decoherence results for non-linear evolution is sorely lacking. Strong decoherence processes at the Plank scale have been proposed by  Hansson\cite{hans0003083} but we make no specific use of this proposal.

 For nonlinear temporal evolution the paradigmatic argument for causality violation is as follows: perform a measurement of an observable \(A\) at time instant \(t_0=0\), allow for a nonlinear evolution for a time interval \(t\), and then perform a measurement of an observable \(B\) in a region space-like separated to the first measuring event. Under nonlinear evolution the expected value of the second measurement will depend on the particular observable \(A\) used and so by switching to a different observable \(A'\) one can communicate across a space-like interval. 

Start with a pure initial state \(\ket\phi\), and  
assume that the first measurement of an observable \(A\) takes place at time \(t=0\). Assume a non-degenerate spectrum:  \(A=\sum_\lambda\lambda\ket{\phi^A_\lambda}\bra{\phi^A_\lambda} \), where \(\ket{\phi^A_\lambda}\) is the eigenvector. In what follows sums are to interpreted as integrals when dealing with continuous spectrum. The state following the measurement is a mixture of the \(\ket{\phi^A_\lambda}\) with probability \(|\!\bracket{\phi^A_\lambda}\phi|^2\).
Let now \(E_t\) be the evolution operator from time \(0\) to time \(t\). The expected
value of \(B\) in the mixture resulting from a measurement of \(A\)
followed by evolution for time \(t\) is now:
\begin{equation}\label{eq:EBtA}
{\cal E}(B,t|A)=
\sum_\lambda|\!\bracket{\phi^A_\lambda}\phi|^2
\melement{E_t\phi^A_\lambda}B{E_t\phi^A_\lambda}
\end{equation}
which also defines the symbol on the left-hand side. Signals can be said
to be present if this depends on the choice of \(A\). It is thus
convenient to define the difference
\begin{equation}
\Delta(B,t|A,A')={\cal E}(B,t|A)-{\cal E}(B,t|A').
\end{equation}
We assume that the measurement event corresponding to 
\(B\) is space-like to the events corresponding to \(A\) and \(A'\).
Expression (\ref{eq:EBtA}) is not very useful as one normally does not have much
knowledge of \(E_t\) since the evolution is generally described through a
differential Schr\"odinger-type equation:
\begin{equation}
i\hbar\partial_t\psi = F\psi
\end{equation}
for some operator \(F\).  We shall make no assumptions of linearity on
\(F\) but we shall assume that the evolution is norm preserving, which,
as was shown in \cite{GGGS:separation}, implies that \(\im
(\psi,F\psi)=0\), a property that is called {\em
norm-hermiticity\/}. It is a difficult mathematical problem
to gain explicit information about \(E_t\). We circumvent it by
expanding into a Taylor series:
\begin{eqnarray*}
{\cal E}(B,t|A) &=& {\cal E}(B,0|A)+ t{\cal E}_1(B|A)+ O(t^2), \\
\Delta(B,t|A,A') &=& \Delta(B,0|A,A')+ t\Delta_1(B|A,A')+ O(t^2).
\end{eqnarray*}
One sees that \({\cal E}(B,0|A)=\melement\phi B\phi\) by the linear quantum mechanical 
no signal theorem, and hence \(\Delta(B,0|A,A')=0\). Since \(B\) is
hermitian we have:
\begin{equation}\label{eq:E1}
{\cal E}_1(B|A)=\frac{2}{\hbar}
\sum_\lambda|\!\bracket{\phi^A_\lambda}\phi|^2\,\im\bracket{B\phi^A_\lambda}{F\phi^A_\lambda}.
\end{equation}
The difference \(t\Delta_1(B|A,A')\) would then determine the signal
amplitude when using a  small delay \(t\).

\section{The DG equation}
 The one-particle DG equation is:
\begin{widetext}
\begin{equation}\label{eq:DG}
i\hbar\partial_t\psi_s = F_s\psi_s = -\frac{\hbar^2}{2m_s}\nabla^2 \psi_s +
iD_s\hbar\left(\nabla^2\psi_s +
\frac{|\nabla\psi_s|^2}{|\psi_s|^2}\psi_s\right)+R_s(\psi)\psi,
\end{equation}
\end{widetext}
where \(s\) labels the particle's species, \(D_s\) is a physical constant and \(R_s(\psi)\) is {\em  real\/} and complex homogeneous: \(R_s(z\psi)=zR_s(\psi)\). We take \(R_s=0\) for simplicity, seeing no motivation for picking any particular form.

For the two particle equation we take 
\begin{equation}i\hbar\partial_t\psi_{ab} =F_{ab}\psi=F^{(1)}_a\psi_{ab}+F^{(2)}_b\psi_{ab}+Q_{ab}\psi
\end{equation}
where \(a\) and \(b\) are species labels, \(F^{(j)}_s\) is the one-particle operator acting on the \(j\)-th particle variables and \(Q_{ab}\) is an operator that vanishes on product functions. This is a canonical construction for separating multi-particle equations as described in \cite{GGGS:separation}. By {\em  separating\/} we mean that uncorrelated two-particle states continue uncorrelated under evolution. This is a non-linear rendition of the notion of lack of interaction and we assume it for simplicity.

It is convenient to
single out the non-real-linear part of \(F_s\) and we define
\begin{equation}
N\phi = \frac{|\nabla\phi|^2}{|\phi|^2}\phi.
\end{equation}

\subsection{The DG equation in the EPR state}
We now see how the
DG equation behaves with respect to the original two-particle EPR state \(\phi_{ab}\)  of zero total momentum.  One performs either
a momentum (\(A=p\)) or a position (\(A'=q\)) measurement on the first particle. Let us see what are the possible contributions to \({\cal E}_1(B|A)\) and \({\cal E}_1(B|A')\) from the various terms in \(F_{ab}\).  Under either one of the measurements, the resulting mixtures 
are of product states \(\ket{\phi^A_\lambda}=\ket{\phi^A_{a\lambda}}\ket{\phi^A_{b\lambda}}\) so the term \(Q_{ab}\) will not contribute, and we assume \(Q_{ab}=0\). Observable \(B\)  acts on particle \(b\) and we have \(\bracket{B\phi^A_{a\lambda}\phi^A_{b\lambda}} {(F_a+F_b)\phi^A_{a\lambda}\phi^A_{b\lambda}}= \bracket{\phi^A_{a\lambda}}{F_a\phi^A_{a\lambda}} \bracket{B\phi^A_{b\lambda}}{\phi^A_{b\lambda}}+ \bracket{\phi^A_{a\lambda}}{\phi^A_{a\lambda}}\bracket{B\phi^A_{b\lambda}}{F_b\phi^A_{b\lambda}}
\). Now the first term is real since \(B\)  is hermitian and \(F_a\) norm-hermitian. The second term is \(\bracket{B\phi^A_{\lambda}}{F_b\phi^A_{\lambda}}\) and so for both \(A\)  and \(A'\) we can use equation (\ref{eq:E1}) with just \(F_b\) in place of \(F_{ab}\), as only the imaginary part of the bracket contributes.

Under momentum measurement the
resulting mixture is of products of momentum eigenstates and  since
\(\nabla^2+N\) vanishes on plane waves one has \(\Delta_1(B|p,q)={\cal E}_1(B|q)\), thus:
\begin{equation}\label{eq:Deltad}
\Delta_1(B|p,q) = 2 D_b\int\re\bracket{B\delta_w}{(\nabla^2+N)\delta_w}\,d\mu(w)
\end{equation}
where \(\delta_w(y)= \delta(y-w)\) is a position eigenstate and \(\mu\)
is the measure defining the mixture. This expression is however not
straightforwardly calculable since a position eigenstate
leads to ill defined forms if we
apply \(N\) to it.
To interpret the
non-liner contribution we use gaussian regularization and so define
\begin{equation}
\delta^{(r)}(y) = \left({r \over \pi}\right)^{n/2}e^{-r y^2}
\end{equation}
as a delta-approximating function which goes over to \(\delta(y)\) as \(r
\to \infty\). Here \(n\) is the dimension of  space. We find then that
\(N(\delta^{(r)})(y)= 4r^2y^2\delta^{(r)}(y)\).
To calculate the asymptotics of this  distribution when \(r \to \infty\), let \(f\) be a Schwartz
test-function and consider
 \begin{eqnarray}\nonumber \lefteqn{4r^2\int y^2 \delta^{(r)}(y)f(y)\,
d^ny =} \\ && =4 \pi^{-n/2} r^{2 + n/2}\int y^2 e^{-r
y^2}f(y)\, d^ny.
\end{eqnarray}
We will make use of the formula
\begin{equation}\label{eq:gaussm}
\int ||y||^p e^{-r
y^2} \, d^ny = {\sigma_n \over 2}\Gamma({n+p\over 2})r^{-{n+p \over
2}}
\end{equation}
which readily follows from 
\[\int_0^\infty e^{-y^q}y^p\, dy =
{1\over q}\Gamma({p+1 \over q}).
\]
 Here \(\sigma_n =
2\pi^{n/2}/\Gamma(n/2)\) is the surface measure
of the unit \(n\)-ball. We see therefore that only for \(p \geq 4\) does
(\ref{eq:gaussm}) overcome  the \(r^{2+  n/2}\)  growth  of  the
coefficient.
We now write
\begin{eqnarray}\nonumber\lefteqn{f(y) = f(0)+  y\cdot (\nabla f)(0) +
{1 \over 2}\sum_{ij}y_iy_jh_{ij}\, +}\\ && + \sum_{ijk}y_iy_jy_kt_{ijk} +
\sum_{ijkl}y_iy_jy_ky_lQ_{ijkl}(y)
\end{eqnarray}
where the \(h_{ij}\)
are the matrix elements of \((Hf)(0)\), the hessian of \(f\) at the origin,
\(t_{ijk}\) a tensor of third derivatives of \(f\) at the origin, and
\(Q_{ijkl}(y)\) are bounded functions of \(y\). If we now consider the
integral \(\int y^2e^{-r y^2}f(y)\, d^ny\) then we see that the
contribution from the fifth term in the expansion for \(f(y)\)
can be bounded by a multiple of \(\int ||y||^6 e^{-r y^2}\, d^ny\) and
is thus of order \(r^{-(n+6) / 2}\). The contribution from the
second and fourth terms vanish since the integrands are odd functions. The
contribution  from  the  first term is  \({n  \over  2}  \pi^{n/2}
r^{-1 -
n/2}f(0)\). To calculate the contribution from the hessian, let us choose
coordinate axes for which the hessian is
diagonal and so we have to calculate \(\int y^2 (h_1y_1^2 + \cdots +
h_ny_n^2) e^{-r y^2}\, d^ny,\) where the \(h_i\) are the eigenvalues.
One type of term is \(\int h_jy_i^2y_j^2 e^{-r y^2} \, d^ny\) with \(i
\neq j\) which
results in \({1 \over 4}\pi^{n/2} r^{-(2+n/2)}h_j,\) and the other
type is \(\int h_jy_j^4 e^{-r y^2} \, d^ny\) which results in \({3 \over
4} \pi^{n/2} r^{-(2+n/2)}h_j\).
Combining all these results we have
\begin{eqnarray*}\lefteqn{\int N(\delta^{(r)})(y)f(y) \, d^ny
=}\\ &&=   2n r f(0) + ({n\over 2} + 1){\rm Tr}((Hf)(0)) + O(r^{-1}),
\end{eqnarray*}
which means that
\begin{equation}
N(\delta^{(r)}) = 2n r \delta + ({n \over 2}+1) \nabla^2 \delta +
O(r^{-1}).
\end{equation}
A similar asymptotic analysis for
\(\delta^{(r)}\)
itself gives \(\delta^{(r)} = \delta + O(r^{-1})\).

Let us now use the Gaussian regularization in (\ref{eq:Deltad}). We
assume that \(B\delta_w\) is well defined and so \(B\delta^{(r)}_w =
B\delta_w +O(r^{-1})\). One
has \(\bracket{B\delta^{(r)}_w}{(\nabla^2+N)\delta^{(r)}_w}= 2rn\melement{\delta_w}B{\delta_w}+
O(1)\) and so
\begin{eqnarray}\nonumber
\Delta_1(B|p,q) &=&
4rn. D_b
\int\melement{\delta_w}B{\delta_w}\,d\mu(w) + O(1) \\ \label{nlrate}
&=&   4rn D_b\melement{\phi}B\phi + O(1).
\end{eqnarray}
We see
therefore that for large enough \(r\), the first term dominates and
any observable whose expectation does not vanish in the initial state
can detect a nonlocal effect that is larger with increased localization.

Separating non-linear evolutions admit additional nonlinear terms first introduced by  Bialynicki-Birula and Mycielski  (\(M\psi=p\ln|\psi|\,\psi\))\cite{bi-bi-my}, and Kostin  (\(K\psi=iq \ln(\psi/\bar\psi)\,\psi\))\cite{kost}. Here \(p\) and \(q\) are real universal constants with the dimension of energy. One has \(\displaystyle M(\delta^{(r)})(y)=\left(\frac n2\ln\frac r\pi-ry^2\right)\delta^{(r)}\) and \(K(\delta^{(r)})(y)=0\). The first expression has asymptotic form \(\displaystyle\frac n2 \ln r\,\delta+O(1)\), and though we have a logarithmic growth here, the \(O(\ln r)\) term is real and will not contribute to (\ref{eq:E1}). The same argument rules out any contribution from the term \(R_s(\psi)\psi\) on the right-hand side of (\ref{eq:DG}) if its asymptotic form on the gaussian is of the form \(\rho(r)\delta+O(1)\). It may be significant that the amplification effect occurs precisely for the diffeomorphism motivated nonlinearity. The amplification seems to be due to two aspects, the singular nature of the nonlinear term when applied to highly localized state, and the fact that the coefficient of \((\nabla^2\psi+N(\psi))\) in the DG equation is {\em  imaginary\/} which adds a diffusion process to the evolution (and breaks time-inversion invariance). A final analysis must await a proper study of nonlinear decoherence and a better understanding of its relation to the diffeomorphism group.

To give  (\ref{nlrate}) more meaning we should compare this rate to one given by a linearly dominated process. As a first approximation we just give a simple  order of magnitude estimate through dimensional analysis. The ratio of nonlinear to the linear coefficient in the DG equation is \(\displaystyle \nu=\frac{D_b}{\hbar/2m}\). 
The rate of change of the expectation value \(\melement\phi{B}\phi\) in a state in which this is dominated by the linear kinetic energy term with a typical one-particle wave-length \(L\)  would be of order \(\displaystyle \frac{\hbar}{mL^2}\melement\phi{B}\phi\). The ratio of the rate given by (\ref{nlrate}) to this is then approximately \(\nu rL^2\).
This expression shows a curious trade-off between short and long wavelength. One needs a high-energy process such as localization to a Plank length to exhibit a nonlocal effect, but the effect is relatively more effective upon a long wavelength state. 
Assume now the localization is to one Planck length \(L_p\), that is \(r=1/L_p^2\). Experimentally \(\nu\) is bounded above by about \(10^{-20}\). At this value, non-linear nonlocal effects would be significant for \(L\) above \(10^{10}L_p\approx 10^{-23}\,{\rm cm.}\) which is likely problematic.  Taking \(L\) of order of the present Hubble radius (\(\approx10^{30}\,{\rm cm.}\)) a value of \(\nu\approx 10^{-126}\)  would guarantee that much shorter wave lengths would not be affected by nonlocal effects. This suppression factor is coincidentally similar to the ratio of the observed density of dark energy   to the quantum field theory predictions of the vacuum energy density. Regardless of the value of the suppression factor \(\nu\), it would provide a fundamental distance scale in nonlinear quantum gravity: \(L_P/\sqrt\nu\). Processes on this scale would be influenced nonlocally by Planck scale processes.  The existence of a third scale (besides \(c\) and \(L_p\)) was recently suggested by Kowalski-Glikman and Smolin\cite{KG.Smo} who estimate it at about \(10^{60}L_p\).  This new scale has implications for dynamics of space-time singularities, in particular for inflationary cosmology, dark energy, and black holes. We shall take up some of these themes in a forthcoming publication.

\subsection*{Acknowledgements}  Part of this research was started during  my sabbatical visit to the Mathematics Department of Rutgers University in 1992-3, for whose hospitality I am grateful. I also wish to thank Professor Gerald Goldin for interesting discussions concerning nonlinear quantum mechanics. This research received partial financial support from the Conselho Nacional de Desenvolvimento Cient\'{\i}fico e Tecnol\'ogico (CNPq), and from the Funda\c{c}\~ao Carlos Chagas Filho de Amparo \`a Pesquisa do Estado do Rio de Janeiro  (FAPERJ).


\begin{thebibliography}{99}
\bibitem{polcal}N.~Gisin,   Helv. Phys. Acta  \textbf{62}, 363 (1989);  M.~Czachor, Found. Phys. Lett. \textbf{4}, 351 (1991);  J.~Polchinski,  
Phys. Rev. Lett. \textbf{66}, 397 (1991);  G.~Svetlichny,
 Found. Phys. \textbf{28}, 131 (1998);  W.~Luecke,
``Nonlocality in Nonlinear Quantum Mechanics",
quant-ph/9904016.

\bibitem{bonaal}M.~Czachor,  Phys. Rev. A~\textbf{53}, 1310 (1996); \textbf{57}, 4122 (1998);  M.~Czachor and ~M.~Kuna, {\sl ibid.} \textbf{58}, 128 (1998);  G.~Svetlichny, in {\sl Proceedings of the Second International Conference ``Symmetry in Nonlinear Mathematical Physics. Memorial Prof. W.~Fushchych Conference"}, edited by  M.~Shkil, A.~Nikitin, and V.~Boyko (Mathematics Institute, National Academy of Sciences of Ukraine, 1997), Vol. 2, p. 262 (proceedings are available online at the Mathematics Institute site: www.imath.kiev.ua/en/);  D.~S.~Abrams and  S.~Lloyd,  Phys. Rev. Lett.
\textbf{81}, 3992 (1998);  P.~Bona, 
``Geometric Formulation of Nonlinear Quantum Mechanics for Density Matrices",
quant-ph/9910011;  M.~Czachor,
 Int. J. Theor. Phys. \textbf{38}, 475 (1999);  S.~Gheorghiu-Svirschevski,
``A General Framework for Nonlinear Quantum Dynamics", quant-ph/0207042;  A.~Kent,
``Nonlinearity without Superluminality", quant-ph/0204106;  M.~Czachor and  H.-D.~Doebner,   Phys. Lett. A~\textbf{301}, 139 (2002).

\bibitem{svetal} G.~Svetlichny,   Journal of Nonlinear Mathematical Physics \textbf{2}, 2  (1995);  G.~Svetlichny, J. Math. Phys. \textbf{45}, 959 (2004).

\bibitem{nlexp1al} J.~J.~Bollinger, D.~J.~Heinzen, W.~M.~Itano, S.~L.~Gilbert, D.~J.~Wineland,  Phys. Rev. Lett. \textbf{63} 1031 (1989);
 R.~L.~Walsworth,  I.~F.~Silvera,  E.~M.~Mattison and  R.~F.~C.~Vessot, {\sl ibid.}\textbf{64} 2599 (1990);
 T.~E.~Chupp and R.~J.~Hoare {\sl ibid.} \textbf{64} 2261 (1990);
P.~K.~Majumder, B.~J.~Venema, S.~K.~Lamoreaux, B.~R.~Heckel, E.~N.~Fortson, {\sl ibid.} \textbf{65} 2931 (1990);
 F.~Benatti and  R.~Floreanini,  Phys. Lett. B~\textbf{389} (1996) 100;
{\sl ibid.} \textbf{451} 422 (1999).


\bibitem{do-gopla162397} H.-D.~Doebner and  G.~A.~Goldin,  Phys. Lett.
 A~\textbf{162}, 397 (1992). 

\bibitem{GGGS:separation} G.~A.~Goldin and  G.~Svetlichny,  {\sl J. Math. Phys.} \textbf{35}, 3322 (1994).

\bibitem{hans0003083} J.~Hansson,
``Nonlinear gauge interactions - A solution to the `measurement problem' in quantum  mechanics?",
quant-ph/0003083.

\bibitem{bi-bi-my} I.~Bialynicki-Birula  and  J.~Mycielski,  Ann. Phys. (N.Y.) \textbf{100}, 62  (1976).

\bibitem{kost} M.~D.~Kostin,  J. Chem. Phys. \textbf{57}, 3589 (1972).

\bibitem{KG.Smo} J.~Kowalski-Glikman and L.~Smolin,  Phys. Rev.  D~\textbf{70}, 065020 (2004).
\end{thebibliography}
\end{document}